\documentclass[aps,prl,tightenlines,twocolumn,groupedaddress,showpacs]{revtex4}
\usepackage{amssymb}
\usepackage{color,graphicx}
\usepackage{lmodern}
\usepackage{amsmath}
\usepackage{amsbsy}
\usepackage{amsthm}
\usepackage{float}
\usepackage{bbm}
\usepackage{bm}
\usepackage{epsfig}

\begin{document}

\title{Emergence of Molecular Chirality due to Chiral Interactions in a\\Biological Environment}

\author{Arash Tirandaz}
\email[]{tirandaz@ch.sharif.edu}
\author{Farhad Taher Ghahramani}
\email[]{farhadtq@ch.sharif.edu}
\author{Afshin Shafiee}
\email[Corresponding Author:~]{shafiee@sharif.edu}
\affiliation{Research Group On Foundations of Quantum Theory and Information,
Department of Chemistry, Sharif University of Technology
P.O.Box 11365-9516, Tehran, Iran}
\begin{abstract}
We explore the interplay between tunneling process and chiral interactions in the discrimination of chiral states for an ensemble of molecules in a biological environment. Each molecule is described by an asymmetric double-well potential and the environment is modeled as a bath of harmonic oscillators. We carefully analyze different time-scales appearing in the resulting master equation at both weak- and strong-coupling limits. The corresponding results are accompanied by a set of coupled differential equations characterizing optical activity of the molecules. We show that, at the weak-coupling limit, chiral interactions prohibit the coherent racemization induced by decoherence effects and thus preserve the initial chiral state. At the strong-coupling limit, considering the memory effects of the environment, the Markovian behavior is retrieved at long times.
 \end{abstract}
\pacs{33.55.+b, 33.80.-b, 87.10.-e, 87.15.B-}
\maketitle

\section{1. Introduction}
A long-standing problem in biology is the origin of biomolecular homochirality, i.e., the observation of a natural preference for L-proteins and D-sugars in life forms~\cite{TeK,Mas,Pal,Kla}. The problem can to some extent be broken down into two sub-problems: why chiral configurations of a chiral molecule are stable ({\it stabilization problem}), and further how to explain the observed preference for a particular chiral configuration of a biological chiral molecule ({\it discrimination problem}). The stabilization problem, more explicitly, refers to the fact that although the states associated to chiral configurations are not stationary states, they are stable for a long time. With some appropriate adiabatic approximations, Hund formulated this problem in a symmetric double-well potential, in which the molecule can bounce between two chiral states -localized in two minima of the potential- by quantum tunneling through the inversion barrier~\cite{Hun}. If the barrier is high enough to prevent the tunneling process, the molecule remains in its initial chiral state. Accordingly once produced, some chiral molecules are stable for a remarkably long time. Upon a quantitative analysis, however, Hund's explanation seems rather insufficient for molecules with low inversion barrier~\cite{Mas1,Jan}.\\
\indent The discrimination problem is addressed directly by considering an asymmetric double-well potential, realized by incorporating chiral interactions into the dynamics. A chiral interaction in general transforms as a pseudo-scalar (i.e., a number that changes sign under parity)~\cite{Baro}. If chiral interactions are strong enough to overcome tunneling process, they can confine the molecule in the chiral state corresponding to the deeper well. The most discussed interactions in this context are parity-violating electro-weak interactions~\cite{Let,Har,Qua,Wes,Dar}. Most theoretical calculations based on these interactions confirm the preference for L-proteins and D-sugars~\cite{Mac}. However, the induced configurational bias is very small to be able to recover the chiral discrimination in a macroscopic scale~\cite{Mas2}. Nevertheless, the external chiral effects (including the dispersion interactions between chiral molecules~\cite{Sto} and interaction with the circularly-polarized light~\cite{Ava}) can be incorporated to amplify these small effects. The problem is that the induced chiral bias is demolished by the inexorable non-linear mechanisms induced by the environment. Here, the challenges are quantifying these mechanisms and introducing them into the dynamics.\\
\indent Many different approaches have been developed to examine the environmental effects. The most cited approaches are the mean-field and decoherence theories. The mean-field theory envisages the effect of the environment as an effective potential added to the Schr\"{o}dinger dynamics of the system~\cite{Bre}. The resulting non-linear dynamics can lead to more stabilized chiral states~\cite{Var,Jon,Gre}. Recently, this model is extended to include chiral interactions in the Langevin formalism of open systems~\cite{Bar1,Bar2,Bar3,Bar4,Bar5}. The decoherence theory describes the environment-induced, dynamical destruction of quantum coherence, which leads to a selection of a distinguished set of system's states~\cite{Giu,Sch}. The most ubiquitous model of decoherence is scattering model or more conveniently collisional decoherence~\cite{Joo,Dio,Adl,Vac}. It was suggested that the chiral states are stabilized due to the collisions with the environment, inducing the indirect position-measurement on the molecule~\cite{Hst1,Hst2,Hst3}. Similar decoherence models also tackled the stabilization problem by coupling the system with photons~\cite{Pfe,Gha} and with phonons~\cite{Fai}. Recently, Trost and Hornberger examined the collisional decoherence of chiral states by dipole-quadruple intermolecular chiral interaction~\cite{Tro}. Bahrami and Shafiee explored the role of fundamental electro-weak interactions on the collisional decoherence of chiral states using the linearized quantum Boltzmann equation~\cite{Bah1}. The collisional decoherence accounts for two-body collisions, appropriate to represent a dilute environment. In a condensed phase, more appropriate to represent a biological environment, since the medium is always present, the idea of a collision loses its meaning. The simplest representation of a condensed phase is a collection of harmonic oscillators. When the oscillators couple linearly to the two-level system, the result is the Spin-Boson model studied extensively in the literature, particularly by Leggett and co-workers~\cite{Leg}. The applications of two-level systems to quantum computation, experiments on macroscopic quantum coherence in SQUIDs, and electron transfer reactions have led to additional interest in the Spin-Boson model (c.f.~\cite{Sch} for a rather complete analysis). Recently, this model is used to analyze the role of quantum decoherence in biological systems~\cite{Gil,Pac,Fls,Hue,Shi,Lei,Zha}. The perturbative treatment of the model in the weak- and strong-coupling limits resulted in the Born-Markov~\cite{Che,Wil} and Nakajima-Zwanzig master equations~\cite{Nak,Zwa}. The path integral method was also used to explore a driven Spin-Boson model~\cite{Gri}. A basic application of this model to chiral molecules is found in the pioneering works of Harris and Silbey~\cite{Har1,Har2,Har3}. They showed that coupling of the molecule to a condensed phase results in the renormalization of the tunneling matrix element. So far, the Spin-Boson model is applied to the case of chiral molecules mostly at the tunneling-dominant limit~\cite{Har3}. Here, we solve the general Spin-Boson model to examine the interplay between tunneling process and chiral interactions in the discrimination of chiral states of an ensemble of chiral molecules in a biological environment. \\
\indent The paper is organized as follows. In the next section, we describe the general Spin-Boson model. In the third section, after a brief introduction on the decoherence theory, we derive the explicit form of the master equations for weak- and strong-coupling limits of the model. The resulting master equations are solved through a set of coupled differential equations to re-examine the chiral discrimination problem by analysing the optical activity of the ensemble of molecules.
\section{2. General Spin-Boson Model}
An ensemble of chiral molecules in interaction with a harmonic bath is characterized by the total Hamiltonian
\begin{equation}\label{E1}
\hat H_{tot}=\hat H+\hat H_{\varepsilon}+\hat H_{int}
\end{equation}
where $\hat H$ and $\hat H_{\varepsilon}$ are the self-Hamiltonian of the molecules and environment, and $\hat H_{int}$ is the interaction Hamiltonian. Each chiral molecule in the ensemble can occur as two identical pair configurations through the inversion at the molecule's center of mass by a long-amplitude vibration known as contortional vibration~\cite{Tow,Her}. This vibration is effectively described by the motion of a particle in an asymmetric double-well potential~\cite{Wei,Bah2}. The minima, associated to equilibrium positions of two chiral configurations, are separated by barrier $V_{\mbox{\tiny${\mbox{\tiny$\circ$}}$}}$. The biased energy, $\omega_{z}$, known as the tilt, is considered as a measure of chiral interactions. In the limit ${{V}_{\mbox{\tiny${\mbox{\tiny$\circ$}}$}}}\gg \hbar{\omega}_{\mbox{\tiny${\mbox{\tiny$\circ$}}$}}\gg k_{\mbox{\tiny$B$}}T$ (${\omega}_{\mbox{\tiny${\mbox{\tiny$\circ$}}$}}$ is the vibration frequency in each well), the molecular states are effectively confined in the two-dimensional Hilbert space spanned by two chiral states. For most chiral molecules this limit holds up to the room temperature~\cite{Tow,Her}. Accordingly, the corresponding effective Hamiltonian for an ensemble with $N$ molecules in the chiral basis can be written as~\cite{Leg}
\begin{equation}
\label{E2}
\hat H=\sum_{i=1}^{N}\big[-\omega_{z}\hat S_{z}-\delta\hat S_{x}\big]_{i}
\end{equation}
where $\delta$ is the tunneling frequency, and $\hat S_{i}$ ($i=x,y,z$) is the $i$-th component of spin Pauli operator. The magnitude of tunneling frequency ranges from the inverse of the lifetime of the universe to thousands of hertz. The tilt introduced into the Hamiltonian slightly stabilizes the right-handed configuration which is the case for D-sugers~\cite{Mac}. The ensemble of molecules is subjected to a condensed bath, modelled as a collection of harmonic oscillators with the Hamiltonian
\begin{equation}\label{E3}
\hat H_{\varepsilon}=\sum_{i}\Big (\frac{1}{2m_{i}}\hat p_{i}^{2}+\frac{1}{2}m_{i}\omega_{i}^{2}\hat q_{i}^{2}\Big)
\end{equation}
The $i$-th harmonic oscillator in the bath is described by its natural frequency, $\omega_{i}$, mass, $m_{i}$, and position and momentum operators, $\hat q_{i}$ and $\hat p_{i}$, respectively. The interaction Hamiltonian is defined as
\begin{equation}\label{E4}
\hat H_{int}=\hat S_{z}\otimes\sum_{i}c_{i}\hat q_{i}
\end{equation}
which describes the correlation between the position of each chiral molecule and the position, $\hat q_{i}$, of $i$-th harmonic oscillator in the environment, with coupling strengths $c_{i}$.
\section{3. Decoherence Program}
In the decoherence approach, at $t=0$, the initial state of each molecule in the ensemble is written as a coherent superposition of two chiral states, $|L\rangle$ and $|R\rangle$, and the environment is assumed to be in a neutral state, $|E_{0}\rangle$. Before the interaction, then, the total ensemble-environment state has the product form
\begin{equation}\label{E5}
|\psi_{tot}(0)\rangle=\prod_{i=1}^{N}\big[c_{L}|L\rangle+c_{R}|R\rangle\big]_{i}\otimes|E_{0}\rangle
\end{equation}
After the interaction, the molecular states become entangled with the corresponding states of the environment
\begin{equation}\label{E6}
|\psi_{tot}(t)\rangle=\prod_{i=1}^{N}\big[c_{L}|L\rangle|E_{L}\rangle+c_{R}|R\rangle|E_{R}\rangle\big]_{i}
\end{equation}
The reduced density matrix of the molecules is obtained by tracing over the environment degrees of freedom
\begin{align}\label{E7}
   \rho(t)&=\prod_{i=1}^{N}\big[|c_{L}|^{2}|L\rangle|\langle L|+|c_{R}|^{2}|R\rangle|\langle R| \nonumber \\
   &+c_{L}c_{R}^{\ast}|L\rangle\langle R|\langle E_{L}|E_{R}\rangle+c_{R}c_{L}^{\ast}|R\rangle\langle L|\langle E_{R}|E_{L}\rangle\big]_{i}
\end{align}
If sufficient information is recorded by the environment, the final environmental states $|E_{L}\rangle$ and $|E_{R}\rangle$ will be approximately orthogonal at the rate $\langle E_{L}|E_{R}\rangle\varpropto e^{-t/\tau_{D}}$, in which $\tau_{D}$ denotes the characteristic decoherence time-scale. Therefore, interferences in the reduced density matrix (\ref{E7}) are suppressed by flowing information from the ensemble of molecules to the environment, leading to a mixed state
\begin{equation}\label{E8}
\hat\rho(t)\approx\prod_{i=1}^{N}\big[|c_{L}|^{2}|L\rangle\langle L|+|c_{R}|^{2}|R\rangle\langle R|\big]_{i}
\end{equation}
The molecular states which emerge dynamically are those states with the most robustness against the interaction with the environment. In the other words, they become least entangled with the environment in the course of the evolution and are thus most immune to decoherence. The rates of entanglement of different sets of molecular states with environment states determine which set of molecular states becomes stable.\\
\indent It is often very difficult, if not impossible, to determine the time evolution of the density matrix in an analytic manner. In such cases, one can use approximation schemes that lead to master equations for the evolution of the reduced density matrix of the molecules. The derivation of a quantum master equation is most easily performed in the interaction picture. The time evolution of total density matrix is determined by the interaction-picture Liouville-von Neumann equation~\cite{Sch}
\begin{equation}\label{E9}
\partial_{t}\hat\rho_{tot}^{(I)}(t)=-\frac{i}{\hbar}\big[\hat H_{int}^{(I)}(t),\hat\rho_{tot}^{(I)}(t)\big]
\end{equation}
The subscript "$I$" indicates the operators in the interaction picture. A typical operator $\hat A(t)$ in the interaction picture is defined as $\hat A^{(I)}=e^{\imath\hat H_{\mbox{\tiny${\mbox{\tiny$\circ$}}$}}t/\hbar}\hat A(t)e^{-\imath\hat H_{\mbox{\tiny${\mbox{\tiny$\circ$}}$}}t/\hbar}$, where $\hat H_{\mbox{\tiny${\mbox{\tiny$\circ$}}$}}=\hat H+\hat H_{\varepsilon}$. From now on, we will simplify our notation by omitting this superscript. We can transform (\ref{E9}) into an equation for the reduced density matrix of the molecules by integrating iteratively, and then taking the trace over the environment states,
\begin{equation}\label{E10}
\partial_{t}\hat\rho(t)=-\int_{0}^{t'}dt~Tr_{\varepsilon}\big[\hat H_{int}(t'),[\hat H_{int}(t),\hat\rho_{tot}(t)]\big]
\end{equation}
If we assume that the interaction between the molecules and the environment is sufficiently weak, the density matrix of the molecules-environment combination remains at all times in an approximate product form ($\hat\rho_{tot}(t)\approx\hat\rho(t)\otimes\hat\rho_{\varepsilon}(t)$). Also, because the environment is large in comparison with the size of the ensemble, the temporal change of the environment density matrix can be neglected ($\hat\rho_{\varepsilon}(t)\approx\hat\rho_{\varepsilon}(0))$. This is called the Born approximation. To proceed further, we should specify the strength of the coupling between the system and environment. The dynamics of the Spin-Boson model can be described at the weak- and strong-coupling limits.
\subsection{I. Weak-Coupling Limit}
\begin{figure}[H]
  \centering
  \includegraphics[width=0.4\textwidth]{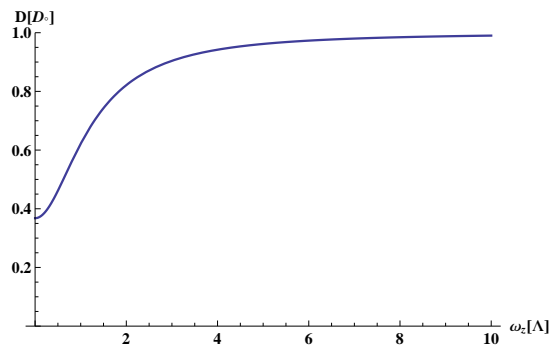}
  \caption{Tilt dependence of the normal decoherence rate $D$ in the units of $D_{\mbox{\tiny${\mbox{\tiny$\circ$}}$}}$ for $\delta=\Lambda$. The tilt, $\omega_{z}$, is in units of $\Lambda$.}
  \label{Fig1}
\end{figure}

At the weak-coupling limit, it is assumed that the environment quickly forget any internal self-correlations that established in the course of the interaction with the molecules. This is called the Markov approximation. Then, one can put the upper limit of the integral in (\ref{E10}) to infinity because the correlation functions are negligible after a definite time. After some mathematics, the Born-Markov master equation is obtained as~\cite{Sch}
\begin{equation}
\label{E11}
\partial_{t}\hat\rho(t)=-\frac{\imath}{\hbar}\big[\hat H,\hat\rho\big]-\pounds_{\rho}
\end{equation}
with
\begin{align}
\label{E12}
  \pounds_{\rho}&=\int_{0}^{\infty} dt \Big\{\nu(t)\big[\hat\sigma_{z},[\hat\sigma_{z}(-t),\hat \rho]\big] \nonumber \\ & \qquad \quad
-\imath\eta(t)\big[\hat\sigma_{z},\{\hat\sigma_{z}(-t),\hat\rho\}\big]\Big\}
\end{align}
The effects of the environment on the dynamics of the reduced density matrix of the molecules are introduced through the noise and dissipation kernels,
\begin{align}
\label{E13}
  \nu(t)&=\int_{0}^{\infty} d\omega~J(\omega)\coth{\big(\frac{\hbar\omega}{2k_{\mbox{\tiny$B$}}T}\big)}\cos{(\omega t)} \nonumber \\ \eta(t)&=\int_{0}^{\infty} d\omega~J(\omega)\sin{(\omega t)}
\end{align}
Here, $J(\omega)$ is the spectral density, corresponding to a continuous spectrum of environmental frequencies, $\omega$. It encapsulates the physical properties of the environment. For more convenience, we employ an ohmic spectral density with an exponential cut-off as
\begin{equation}
\label{E14}
J(\omega)=J_{\mbox{\tiny${\mbox{\tiny$\circ$}}$}}\omega e^{-\omega/\Lambda}
\end{equation}
in which $J_{\mbox{\tiny${\mbox{\tiny$\circ$}}$}}$ is a dimensionless measure of the system-environment coupling strength and $\Lambda$ is a high-frequency cut-off. The general Spin-Boson model, defined by the Hamiltonian (\ref{E2}), is complex, both in terms of its mathematical solution and the dependencies of the resulting decoherence dynamics on the relative magnitudes of the parameters. More specifically, the time dependence of $\hat\sigma_{z}$ in the interaction picture can be carried out thorough the heavy calculations of the disentangling theorem~\cite{Pur}. For a more trackable discussion, we pass over the details of the calculation, and present the final result as
\begin{equation}
\label{E15}
\hat\sigma_{z}(-t)=A_{1}(t)\hat\sigma_{z}+A_{2}(t)\hat\sigma_{x}+A_{3}(t)\hat\sigma_{y}
\end{equation}
with
\begin{align}
\label{E16}
 A_{1}(t)&=\frac{\omega^{2}_{z}+\delta^{2}\cos{(kt)}}{k^{2}}\nonumber \\
 A_{2}(t)&=\frac{\omega_{z}\delta}{k^{2}}\big[1-\cos{(kt)}\big]  \nonumber \\
 A_{3}(t)&=\frac{\delta}{k}\sin{(kt)}
\end{align}
with $k^{2}=\omega^{2}_{z}+\delta^{2}$. The first term in (\ref{E15}) denotes the population difference for an isolated ensemble of chiral molecules, as previously reported by Harris and Stodolsky~\cite{Har}, and recently by Bargue\~{n}o and co-workers~\cite{Bar1} in the Langevin formalism. If we have a symmetric double-well, i.e., $\omega_{z}=0$, the well-known symmetric tunneling oscillations are recovered. According to the explicit expression of $A_{1}(t)$ in (\ref{E16}), even if the magnitude of chiral interactions is small, they may swamp the effect of tunneling.\\
\indent After inserting (\ref{E15}) in (\ref{E12}), the non-linear part of the Born-Markov master equation reads as
\begin{align}
\label{E17}
  \pounds_{\rho} &=D\big[\hat\sigma_{z},\big[\hat\sigma_{z},\hat\rho\big]\big]  \nonumber \\
  & \quad +f[\hat\sigma_{z},\big[\hat\sigma_{x},\hat\rho\big]\big]+f'\big[\hat\sigma_{z},\big[\hat\sigma_{y},\hat\rho\big]\big]\nonumber \\
  & \quad -\imath\gamma\big[\hat\sigma_{z},\big\{\hat\sigma_{x},\hat\rho\big\}\big]-\imath\gamma'\big[\hat\sigma_{z},\big\{\hat\sigma_{y},\hat\rho\big\}\big]\big)
\end{align}
with
\begin{align}
\label{E18}
  D&=\int_{0}^{\infty} dt~\nu(t)A_{1}(t)  \nonumber \\
  f&=\int_{0}^{\infty} dt~\nu(t)A_{2}(t), \quad f'=\int_{0}^{\infty} dt~\nu(t)A_{3}(t)  \nonumber \\
  \gamma &=\int_{0}^{\infty} dt~\eta(t)A_{2}(t), \quad \gamma'=\int_{0}^{\infty} dt~\eta(t)A_{3}(t)
\end{align}
This is the complete form of the Born-Markov master equation for the Spin-Boson model. Each coefficient in the master equation (\ref{E17}) carries a particular physical interpretation, resulted from its role in the master equation.\\
\indent The first term in (\ref{E17}) describes the direct monitoring of the molecules position by the environmental particles. Thus it describes normal decoherence, or more conveniently dephasing, at a rate given by
\begin{equation}
\label{E19}
D=\frac{\pi J_{\mbox{\tiny${\mbox{\tiny$\circ$}}$}}k_{\mbox{\tiny$B$}}T\omega^{2}_{z}}{\hbar k^{2}}+\frac{\pi J_{\mbox{\tiny${\mbox{\tiny$\circ$}}$}}\delta^{2}}{2k}e^{-\frac{k}{2\Lambda}}\coth{\big(\frac{\hbar k}{2k_{\mbox{\tiny$B$}}T}\big)}
\end{equation}
At the high-temperature limit, it is assumed that the thermal energy, $k_{\mbox{\tiny$B$}}T$, of the environment is much greater than the natural energy of the molecules $\hbar k$ and the environment cut-off energy, $\hbar\Lambda$~\cite{Cal}. At this limit, we can approximate~$\coth{\big(\frac{\hbar k}{2k_{\mbox{\tiny$B$}}T}\big)}\approx\frac{2k_{\mbox{\tiny$B$}}T}{\hbar k}$. Therefore, the decoherence rate is reduced to
\begin{equation}
\label{E20}
D\simeq D_{\mbox{\tiny${\mbox{\tiny$\circ$}}$}}\Big(\frac{\omega_{z}^{2}+\delta^{2} e^{-\frac{k}{2\Lambda}}}{k^{2}}\Big)
\end{equation}
with $D_{\mbox{\tiny${\mbox{\tiny$\circ$}}$}}=\pi J_{\mbox{\tiny${\mbox{\tiny$\circ$}}$}}k_{\mbox{\tiny$B$}}T/\hbar$. FIG.~\ref{Fig1} portrays the behavior of the decoherence rate $D$ against the tilt. The plot clearly shows that decoherence effects resulted from the symmetric double-well (i.e., $\omega_{z}=0$), where the dynamics is governed by tunneling process, stabilizing the chiral stats at the rate $D=D_{\mbox{\tiny${\mbox{\tiny$\circ$}}$}}e^{-\frac{\delta}{\Lambda}}$ ({\it stabilization problem}). As mentioned, for an isolated chiral molecule, introducing chiral interactions into the dynamics through the tilt, stabilizes the corresponding chiral state. For a chiral molecule in interaction with the environment, this means that the environment affects two chiral states differently. Since the decoherence rate increases with the tilt, decoherence effects stabilize this particular chiral state more than the other one ({\it discrimination problem}).
\noindent The second and third terms in (\ref{E17}), known as anomalous decoherence terms, describe the position monitoring of the molecules by another variables of the bath. The anomalous decoherence of $\hat\sigma_{x}$ occurs at the rate
\begin{equation}
\label{E21}
  f=\frac{\pi J_{\mbox{\tiny${\mbox{\tiny$\circ$}}$}}\omega_{z}\delta}{k}\Big[\frac{k_{\mbox{\tiny$B$}}T}{\hbar k}-\frac{1}{2}e^{-\frac{k}{2\Lambda}}\coth{\big(\frac{\hbar k}{2k_{\mbox{\tiny$B$}}T}\big)}\Big]
\end{equation}
\noindent and the analogues anomalous rate $f'$ is obtained as
\begin{equation}
\label{E22}
f'=\frac{\sqrt{\pi} J_{\mbox{\tiny${\mbox{\tiny$\circ$}}$}}k_{\mbox{\tiny$B$}}T\delta}{2\hbar\omega_{z}}\Bigg\{G_{1,3}^{2,1}\left( \big(\frac{k}{2\Lambda}\big)^{2} \  \Bigg\vert \  {\frac{1}{2} \atop \frac{1}{2},\frac{1}{2},0} \right)\Bigg\}
\end{equation}
where $G$ is the Meijer G-function~\cite{Mat}. The tilt dependence of the anomalous rates exhibits a decay behaviour. This is because the augmentation of direct monitoring of molecule position with the tilt suppresses its indirect monitoring.\\
\indent The temperature dependence of decoherence rates are approximately linear. In the other words, decoherence becomes stronger as the temperature of the environment is raised. This increase is due to the fact that, as the temperature is raised, excited energy levels will be occupied in each harmonic oscillator increasingly, and thus the characteristic wavelengths present in the bath will decrease. This means that the bath will be able to better resolve the position of the molecules, leading to stronger decoherence of superpositions of well-separated positions.\\
\indent The two last terms of the master equation (\ref{E17}) describe $\hat\sigma_{x}$ and $\hat\sigma_{y}$ damping-and thus dissipation-due to the interaction with the environment. Note that dissipation coefficients are independent of the reservoir temperature. The damping of $\hat\sigma_{x}$ occurs at a rate given by
\begin{equation}
\label{E23}
  \gamma =\gamma_{\mbox{\tiny${\mbox{\tiny$\circ$}}$}}\frac{\omega_{z}\delta}{k^{2}}\Big[(1-\sqrt{\pi}G_{1,3}^{2,1}\left( \big(\frac{k}{2\Lambda}\big)^{2} \  \Bigg\vert \  {2 \atop 2,2,\frac{2}{3}} \right))\Big]
\end{equation}
with $\gamma_{\mbox{\tiny${\mbox{\tiny$\circ$}}$}}=J_{\mbox{\tiny${\mbox{\tiny$\circ$}}$}}\Lambda$. The analogues dissipation term occurs at the rate
\begin{equation}
\label{E24}
\gamma'=\gamma_{\mbox{\tiny${\mbox{\tiny$\circ$}}$}}\frac{\pi\delta}{2\Lambda}e^{-\frac{k}{\Lambda}}
\end{equation}
The tilt dependence of the dissipation rates exhibits a decay behavior. Dissipation causes the system ultimately to reach thermal equilibrium. The characteristic rate on which this happens is typically referred to as the relaxation rate. This rate is directly dependent on the strength of the interaction between the system and environment. For the system under study, increasing the strength of chiral interactions through the tilt increases energy difference between two wells. If this energy difference approaches the maximum energy of the environment, cut-off energy, $\hbar\Lambda$, the environmental particles cannot monitor the molecules and then the coupling between molecules and environment becomes weaker, which in turn decreases the relaxation rate.
\subsubsection{Biological Environments}
Now we discuss the implications of the master equation that we developed to describe the interactions between biomolecules and their surroundings. The biomolecule-environment coupling plays an important role in the behavior of biomolecules. It is necessary to have good models for the environment to show how the resulting decoherence will affect the biomolecule's dynamics (for a complete treatment of modelling decoherence in biomolecules, see~\cite{Bot}). An interesting class of models focuses on the interaction between a two-level system, described by Pauli matrices, and its environment. A particular specific form for this interaction, applicable to a diverse range of systems is the Spin-Boson model, which we discussed here. Now, we again use the master equation (\ref{E11}) to describe the behavior of chiral molecules in interaction with a biological environment. To be more explicit, we focus our attention on a chromophore. A chromophore is an optically active part of a protein responsible for its color. A natural environment for the chromophore consists of its protein and surrounding solvent. The chromophore and its condensed environment can be modeled as a two-level system and a bath of harmonic oscillators~\cite{Gil2}. The coupling with the environment is determined by the spectral density $J(\omega)$. The spectral density is obtained from the microscopic details of the model under consideration. The simplest model arises when the chromophore is treated as a point dipole inside a uniform, spherical protein surrounded by a uniform polar solvent. For a Debye solvent and a protein with a static dielectric constant, the spectral density can be written as~\cite{Gil2}
\begin{equation}\label{E25}
J(\omega)=\frac{\alpha\omega}{1+\tau^{2}_{E}\omega^{2}}
\end{equation}
 with
\begin{equation}
\label{E26}
\alpha=\frac{(\Delta\mu)^{2}}{4\pi\epsilon_{\mbox{\tiny${\mbox{\tiny$\circ$}}$}}b^{3}}\frac{6\epsilon_{p}(\epsilon_{s}-\epsilon_{\infty})\tau_{E}}{(2\epsilon_{s}+\epsilon_{p})(2\epsilon_{\infty}+\epsilon_{p})}
\end{equation}
and
\begin{equation}
\label{E27}
\tau_{E}=\frac{2\epsilon_{\infty}+\epsilon_{p}}{2\epsilon_{s}+\epsilon_{p}}\tau_{D}
\end{equation}
where $b$ is the radius of the protein containing the chromophore, $\Delta\mu$  is the difference between the dipole moment of the chromophore in the ground and excited states, $\epsilon_{p}$  is a dielectric constant of the protein environment, $\epsilon_{s}$ and $\epsilon_{\infty}$ are the static and high-frequency dielectric constants of the solvent, and $\tau_{D}$ is the Debye relaxation time of the solvent. For a chromophore in water, we have $\alpha\approx1$ and $\tau_{E}\approx0.5-2.5ps$. The arbitrary choice of the cut-off does not change the structure of the reservoir and thus the exponential cut-off, used in (\ref{E14}), yields the same dynamics. So, we can employ the ohmic version of biological spectral density with $J_{\mbox{\tiny${\mbox{\tiny$\circ$}}$}}=\alpha$ and $\Lambda=\tau_{E}^{-1}$. The previous discussion can be applied to biological chiral molecules by incorporating this allocation. At this point, we can estimate an order of magnitude for the weak-coupling dephasing rate for an ensemble of Adamantanone molecules in an aqueous environment. The high-temperature cut-off frequency, according to the limit $k_{\mbox{\tiny$B$}}T\gg\hbar\Lambda$, for the temperature $100K$ can be estimated as $10^{12}Hz$. The tunneling frequency of Adamantanone is about to $10^{12}Hz$. Taking $\omega_{z}=10^{3}\delta$, the weak-coupling dephasing rate is estimated as $10^{12}Hz$. In the next section, we examine optical activity of the ensemble of Adamantanone molecules in water as a common biological environment.
\subsubsection{Optical Activity}
\begin{figure}[H]
  \centering
  \includegraphics[width=0.45\textwidth]{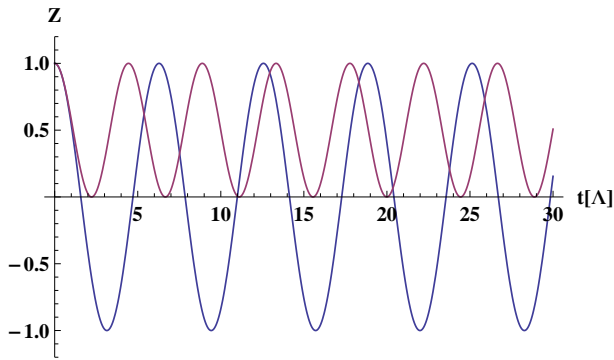}
  \caption{Time evolution of optical activity of an isolated chiral molecule for $\omega_{z}=0$ (blue) and $\omega_{z}=\delta$ (red).}
  \label{Fig2}
\end{figure}

\begin{figure}[H]
  \centering
  \includegraphics[width=0.45\textwidth]{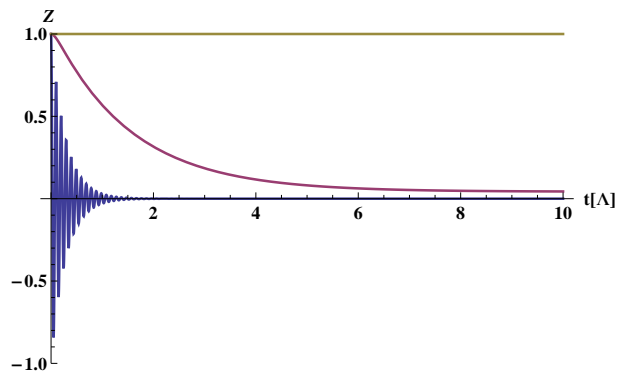}
  \caption{Time evolution of optical activity for an ensemble of Adamantanone molecules in an aqueous environment at the tunneling-dominant limit (blue) for $\omega_{z}=10^{-3}\delta$, interplay limit (red) for $\omega_{z}=\delta$ and localization-dominant limit (green) for $\omega_{z}=10^{3}\delta$. The initial state is the right-handed state and the temperature is fixed at $T=100K$.}
  \label{Fig3}
\end{figure}

\begin{figure}[H]
  \centering
  \includegraphics[width=0.45\textwidth]{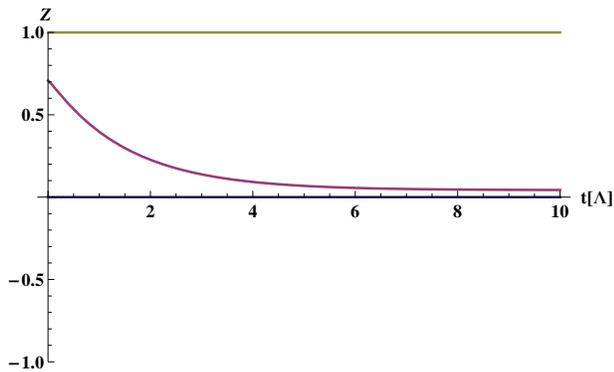}
  \caption{Same as FIG.~\ref{Fig3} for the initial ground state.}
  \label{Fig4}
\end{figure}

\begin{figure}[H]
  \centering
  \includegraphics[width=0.5\textwidth]{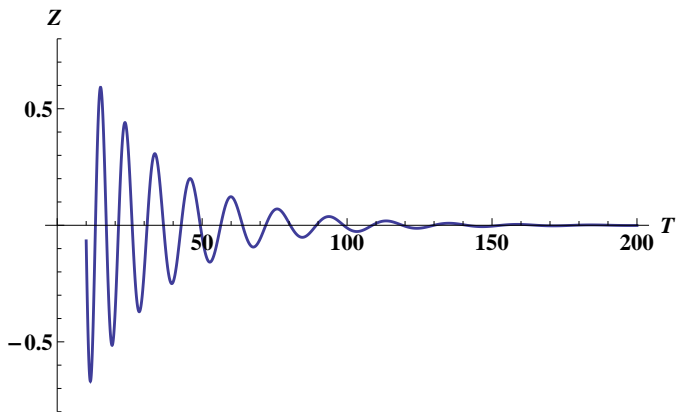}
  \caption{Temperature dependence of optical activity of an ensemble of Adamantanone molecules in aqueous environment at the tunneling-dominant limit and the initial right-handed preparation at $t=10^{-12}s$.}
  \label{Fig5}
\end{figure}

Optical activity is the ability of an ensemble of chiral molecules to show chiroptical properties including optical rotation, circular dichroism and differential Rayleigh and Raman scattering of circularly polarized light. It can be used to determine which chiral configuration is present in the ensemble of molecules. Optical activity is proportional to the population difference, $Z$, between two corresponding chiral states. We begin with isolated chiral molecules. Optical activity for an ensemble of chiral molecules in the absence of environmental interactions is determined by the first expression in (\ref{E16}). The corresponding behavior, blue plot in FIG.~\ref{Fig2}, shows symmetric oscillations between two chiral states. Introducing asymmetry into the unitary dynamics reduces the symmetry of the oscillations, eventually confining them to one well (red plot in FIG.~\ref{Fig2}). Since the ensemble of molecules is isolated, this oscillatory behavior can be interpreted as the quantum signature of optical activity.
\noindent Now, we turn our attention into an ensemble of chiral molecules in interaction with the environmental particles. The coupling of the molecules with the environment has two immediate consequences. First, it turns an ensemble of chiral molecules, containing predominantly one chiral configuration, into a mixture of equal amount of each chiral configuration. This is a relaxation process, known as dephasing racemization, during which optical activity of the ensemble decreases. This type of racemization should be distinguished from the typical thermal racemization. In the former, the incoherent tunneling through the barrier is induced by the environment, while in the later, the molecules are energetic enough to surmount the barrier. Here, we assume that the temperature is sufficiently low, so that the thermal racemization is improbable. Second, it suppresses the quantum characteristics of the molecules by reducing the amplitude of the oscillations. The Born-Markov master equation (\ref{E17}) is conveniently solved through the solution of the time evolution of the elements of the density matrix, determined by the set of differential equations
\begin{align}
\label{E28}
   \partial_{t}X(t)&=-DX(t)+\omega_{z}Y(t)+(f+\gamma)Z(t)+\gamma' \nonumber \\
   \partial_{t}Y(t)&=-\omega_{z}X(t)-DY(t)+(f'+\delta)Z(t)-\gamma \nonumber \\
   \partial_{t}Z(t)&=-\gamma X(t)-\delta Y(t)
\end{align}
with $X(t)=Re(\rho_{LR}(t))$, $Y(t)=Im(\rho_{LR}(t))$ and $Z(t)=\rho_{RR}(t)-\rho_{LL}(t)$ for the Bloch-vector representation~\cite{All}. This is the complete form of Bloch-type equations for the Spin-Boson model. The stability of molecules is analyzed based on the character of the eigenvalues of the Jacobian matrix. The general form of population difference can be written as $Z(t)\sim\sum_{n=1}^{3}A_{n}e^{\lambda_{n}t}\cos{\alpha_{n}t}$, where $A_{n}$ is the amplitude coefficient and $\lambda_{n}+\imath\alpha_{n}$ is $n$th eigenvalue of Jacobian matrix. If all eigenvalues have negative real parts ($\lambda_{n}\leq0$), the molecules are stable. In this case, population difference $Z(t)$ acts like an under-damped oscillator. The competition between damping (exponential term) and oscillation (periodic term) determines the dynamics of the system.\\
\indent Now, we discuss the dynamics of optical activity for different initial preparations. We assume that the initial state of the ensemble of molecules is prepared as
\begin{equation}\label{E29}
\rho_{0}=\left(
\begin{array}{cc}
P_{R} & a-\imath b  \\
a+\imath b & P_{L}
\end{array} \right)
\end{equation}
Here, $P_{R}$ and $P_{L}$ are the probabilities to find the molecule in the right- and left-well, and $a$ and $b$ are the coherences. In the standard preparation, the system is set up at time $t=0$ in a localized eigenstate of $\hat S_{z}$, in particular when the tunneling dynamics is investigated. We then have $P_{R/L}=\pm1$ and $a = b = 0$. Optical activity of the ensemble under environmental interactions can be studied in three limits: tunneling-dominant limit, localization-dominant limit, and interplay limit (the limit at which tunneling and localization can compete with each other). At the tunneling-dominant limit, corresponding to the symmetric double-well, after several symmetric oscillations (resulted from tunneling dynamics), racemization (resulted from the dissipation effects of the environment) occurs, and destroys the initial chiral state (blue plot in FIG.~\ref{Fig3}). If we introduce the chiral interactions into the dynamics, oscillations are gradually suppressed, but racemization still occurs (red plot in FIG.~\ref{Fig3}). At the localization-dominant limit, the chiral interactions prohibit the dissipative effects of the environment and then confine the molecules in the initial chiral state (green plot in FIG.~\ref{Fig3}).
\noindent When the molecules are prepared in the ground state of energy, we have $P_{R}-P_{L}=\omega_{z}/k$, $a =\delta/k$ and $b = 0$. For this initial preparation, different regimes correspond to different initial conditions (FIG.~\ref{Fig4}). At the tunneling-dominant limit, since the amplitude coefficients are zero, there is no dynamics. The dynamics of the molecules at the interplay and localization limits are similar to the corresponding dynamics in the localized initial state.
\noindent The comparison between the localized state and ground state behaviours shows that although the equilibrium states reached asymptotically are independent of the initial state, effects of the initial preparation in the under-damped regime strongly affect the short time dynamics. These results are also obtained by Grifoni and co-workers which studied the driven Spin-Boson model using the path integral method~\cite{Gri}.\\
\indent A decrease in the temperature of the bath reduces the decoherence effects, realized by an augmentation in the oscillatory behavior of optical activity (FIG.~\ref{Fig5}), which is in agreement with the work of Pe\~{n}ate-Rodr\'{i}guez and co-workers~\cite{Bar5} in the Langevin approach.
Our results are also consistent with the results of the mean-field approach. In this context, Vardi employed a Hartree-Fock technique to account for the interaction between each molecule and the mean-field induced by all other molecules of the ensemble~\cite{Var}. This mean-field includes both homochiral (between molecules of the same chirality) and hetrochiral (between molecules of opposite chirality) interactions. When the non-linearity given by the difference between hetrochiral and homochiral interactions is sufficiently large, the population is trapped in one of the wells even if the potential is perfectly symmetric. The origin of this difference can be traced back to the intermolecular chiral interactions. In our study, as it is convenient in the decoherence approach, we describe the dynamics of the isolated ensemble of molecules with a linear dynamics. Therefore, all chiral interactions are included in the tilt. If we introduce the tilt into the dynamics, the oscillations are confined to one well (red plot in FIG.~\ref{Fig2}), which is consistent with the Vardi's result. Recently, Bargue\~{n}o and co-workers extended Vardi's model to include other chiral interactions and dissipative effects of the environment in the Langevin formalism of open systems~\cite{Bar1}. They showed that considering the dissipative effects of the environment, the interplay between the mean-field and external chiral effects leads to the localization, independent of the initial preparation. In our work, localization occurs at the localization-dominant limit, independent of the initial state (green plots in FIG.~\ref{Fig3} and FIG.~\ref{Fig4}). In a more recent work, Gonzalo and Bargue\~{n}o have studied the effect of an external chiral field on the optical activity of chiral molecules in the gas phase~\cite{Bar2}. They showed that decoherence effects always lead to racemization, for small enough chiral external field. In the condensed phase, however, since the environment is always present, the chiral interactions can preserve the initial chiral state.
\subsection{II. Strong-Coupling Limit}
\begin{figure}[H]
  \centering
  \includegraphics[width=0.45\textwidth]{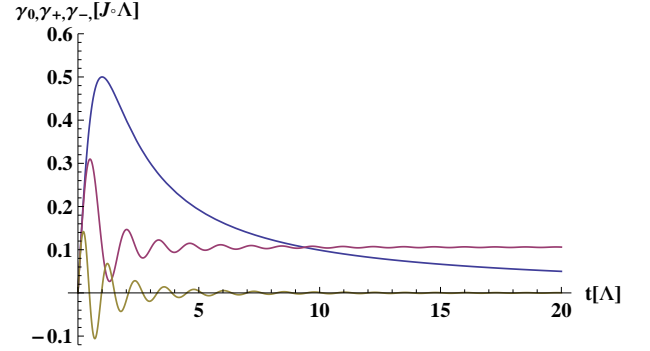}
  \caption{Time dependences of rates $\gamma_{0}(t)$ (blue), $\gamma_{+k}(t)$ (red) and $\gamma_{-k}(t)$ (green) for $k=5\Lambda$ in the units of $J_{\mbox{\tiny${\mbox{\tiny$\circ$}}$}}\Lambda$.}
  \label{Fig6}
\end{figure}

At the strong-coupling limit, which is likely at very low temperatures, since the memory effects of the environment are at work, the dynamics is non-Markovian. The pronounced memory effects in the environment may cause strong dependencies of the evolution of the reduced density matrix on the past history of the system-environment combination. This makes it impossible to describe the reduced dynamics by a time-local differential equation. However, it is often possible to arrive at a non-Markovian but time-local master equations by using the so-called time-convolutionless projection operator technique~\cite{Zwa}. This ansatz provides a perturbative expansion of the system-environment interaction and results in a local time evolution equation for the reduced density matrix of the system.\\
\indent Consider a system which is coupled to an environment. The Hamiltonian of the total system is given by
\begin{equation}\label{E30}
\hat H_{tot}=\hat H_{\mbox{\tiny${\mbox{\tiny$\circ$}}$}}+\alpha\hat H_{int}
\end{equation}
where $\hat H_{\mbox{\tiny${\mbox{\tiny$\circ$}}$}}$ describes the free evolution of the system and environment, and $\alpha$ is an expansion parameter that determines the strength of the coupling. The state of the total system is described by the density matrix, $\hat\rho_{tot}$, which is a solution of the Liouville-von Neumann equation
\begin{equation}\label{E31}
\partial_{t}\hat\rho_{tot}(t)=-\frac{i\alpha}{\hbar}\big[\hat H_{int}(t),\hat\rho_{tot}(t)\big]
\end{equation}
If we integrate over this equation twice and then differentiate with respect to $t$, we obtain the time-convolutionless equation to second order which is the same as (\ref{E10}). Recently, Laine derived a time-convolutionless master equation for the Spin-Boson model~\cite{Lai}. Here, we apply Laine's work to the case of chiral molecules. The interaction Hamiltonian, $\hat H_{int}$, is taken to be a tensor product of operator, $\hat S$, of the molecules and operator, $\hat E$, of the environment; $\hat H_{int}=\hat S\otimes\hat E$. The molecule's operator may be decomposed into its eigenoperators as $\hat S=\sum_{\omega'}\hat S(\omega')$, where $\omega'=\varepsilon'-\varepsilon/\hbar$ with $\varepsilon$ and $\varepsilon'$ being the eigenvalues of the molecular Hamiltonian. For our Spin-Boson model, the eigenoperator of each molecules are defined as
\begin{equation}
\label{E32}
\hat S(0)=-\frac{\omega_{z}}{2k}\hat\sigma_{z},~\hat S(\pm k)=\frac{\delta}{k}\hat\sigma_{\mp}
\end{equation}
where $\hat\sigma_{\mp}$ are ladder operators. To obtain a physically intuitive description of the dynamics, it is convenient to use the secular approximation which consists of replacing the generator of the interaction-picture master equation by its time average. The corresponding master equation can be written as~\cite{Lai}
\begin{equation}\label{E33}
\partial_{t}\rho(t)=-\frac{\imath}{\hbar}\big[\hat H_{LS},\hat\rho\big]-\pounds'_{\rho}
\end{equation}
with
\begin{align}\label{E34}
\pounds'_{\rho}&=\sum_{\omega'}\gamma_{\omega'}(t)\big[\hat S(\omega')\hat \rho \hat S^{\dagger}(\omega')\nonumber \\ &\qquad\quad  -\frac{1}{2}\big\{\hat S^{\dagger}(\omega')\hat S(\omega'),\hat \rho\big\}\big]
\end{align}
where $\hat H_{LS}$ is the Lamb-shift Hamiltonian and the time-dependent rate is defined as
\begin{equation}\label{E35}
   \gamma_{\omega'}(t)=\int_{0}^{t} dt'\int_{0}^{\infty} d\omega J(\omega)\sin{(\omega'-\omega)t'}
\end{equation}
which for ohmic spectral density (\ref{E14}) yields
\begin{align}\label{E36}
  \gamma_{0}(t)&=\frac{J_{\mbox{\tiny${\mbox{\tiny$\circ$}}$}}\Lambda^{2}t}{1+\Lambda^{2}t^{2}} \nonumber \\
   \gamma_{\pm k}(t)&=\frac{J_{\mbox{\tiny${\mbox{\tiny$\circ$}}$}}\Lambda}{1+\Lambda^{2}t^{2}}\big[\Lambda t\cos{(kt)}\mp\sin{(kt)}\big] \nonumber \\ &\quad+J_{\mbox{\tiny${\mbox{\tiny$\circ$}}$}}ke^{\mp\frac{k}{\Lambda}}\Big[\Re{\big[Si(kt+\imath\frac{k}{\Lambda})\big]} \nonumber \\ &\quad\qquad\quad\mp\Im{\big[Ci(kt+\imath\frac{k}{\Lambda})\big]}\pm\frac{\pi}{2}\Big]
\end{align}
where $\Re$ and $\Im$ denote the real and imaginary parts, and $Si$ and $Ci$ are sine- and cosine-integral, respectively. FIG.~\ref{Fig6} depicts the time evolution of the rates. While the rate $\gamma_{0}(t)$ remains always positive, the rates $\gamma_{\pm k}(t)$ have oscillatory behaviour and get temporarily negative values resembling memory effects in the reduced dynamics of the molecules.
\noindent The equations of motion for the elements of the reduced density matrix are given by
\begin{align}\label{E37}
\partial_{t}\rho_{LR}(t)&=-\frac{1}{2k^{2}}\big[\omega_{z}^{2}\gamma_{0}(t)+\delta^{2}\gamma^{+}(t)\big]\rho_{LR}(t)\nonumber \\
\partial_{t}\rho_{LL}(t)&=\frac{\delta^{2}}{4k^{2}}\big[\gamma_{-k}(t)-\gamma^{+}(t)\rho_{LL}(t)\big]
\end{align}
in which we defined $\gamma^{+}(t)=\gamma_{k}(t)+\gamma_{-k}(t)$. The solution of this set of differential equations can be written as
\begin{equation}\label{E38}
\rho(t)=M(t)\rho(0)
\end{equation}
with
\begin{equation}\label{E39}
M(t) =
 \begin{pmatrix}
  g(t) & 0 & 0 & f(t) \\
  0 & e^{-\zeta(t)} & 0 & 0 \\
  0  & 0  & e^{-\zeta(t)} & 0  \\
  1-g(t) & 0 & 0 & 1-f(t)
 \end{pmatrix}
\end{equation}
and,
\begin{align}\label{E40}
   g(t)&=f(t)+e^{-\eta(t)}  \nonumber\\
   f(t)&=e^{-\eta(t)}\xi(t) \nonumber\\
   \eta(t)&=\frac{\delta^{2}}{4k^{2}}\int_{0}^{t} dt~\gamma^{+}(t)\nonumber\\
   \xi(t)&=\frac{\delta^{2}}{4k^{2}}\int_{0}^{t} dt~\gamma_{-k}(t)e^{-\eta(t)}\nonumber\\
   \zeta(t)&=\int_{0}^{t}dt~\Big(\frac{\omega_{z}^{2}}{2k^{2}}\gamma_{0}(t)+\frac{\delta^{2}}{8k^{2}}\gamma^{+}(t)\Big)
\end{align}
The matrix (\ref{E39}) acts on $\rho(0)$ as a column vector consisting of the density matrix elements.\\
\indent Note that because weak- and strong-coupling master equations are derived through different assumptions, their terms cannot be compared one-to-one. Nevertheless, the contribution of dephasing process (indicated by superscript "d") can still be distinguished by the following equation of motion
\begin{equation}\label{E41}
  \partial_{t}\rho^{d}_{LR}(t)=-\frac{\omega_{z}^{2}}{2k^{2}}\gamma_{0}(t)~\rho^{d}_{LR}(t)
\end{equation}
The solution of the dephasing equation is obtained as
\begin{equation}\label{E42}
\rho^{d}_{LR}(t)=\big(1+t^{2}\Lambda^{2}\big)^{-J_{\mbox{\tiny${\mbox{\tiny$\circ$}}$}}\omega_{z}^{2}/4k^{2}}\rho^{d}_{LR}(0)
\end{equation}
The decay rate due to the dephasing process, i.e., the rate at which the value of off-diagonal elements is reduced to $1/e$ times its initial value, is given by
\begin{equation}\label{E43}
D=\Big(e^{\frac{4}{J_{\mbox{\tiny${\mbox{\tiny$\circ$}}$}}}(1+\frac{\delta^{2}}{\omega_{z}^{2}})}-1\Big)^{-1/2}\Lambda
\end{equation}
It can be easily seen that similar to the analogous rate of weak-coupling limit, the strong-coupling dephasing rate increases with the ratio of tilt to tunneling, but at a lower rate. Now, we can estimate the order of magnitude of the strong-coupling-temperature dephasing rate for an ensemble of Adamantanone molecules in an aqueous environment. The cut-off frequency, corresponding to the low-temperature limit ($k_{\mbox{\tiny$B$}}T\ll\hbar\Lambda$), for the temperature $0.1K$ is estimated as $10^{9}Hz$. Taking $\omega_{z}=10^{3}\delta$, corresponding to the localization-dominant limit, the dephasing rate would be $10^{10}Hz$. The lower rate of strong-coupling dephasing process shows that the environmental interactions at low temperatures require more time to suppress the quantum characteristics of the molecules and stabilize the chiral states. This is because the memory effects of the environment at low temperatures reduce the distinguishability of environmental states. Therefore, the rate of leakage of the quantum correlations of the molecules to the environment will be reduced and the molecules become more isolated.
\subsubsection{Optical Activity}
\begin{figure}[H]
  \centering
  \includegraphics[width=0.4\textwidth]{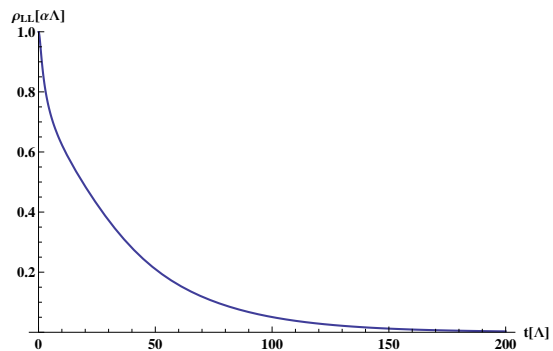}
  \caption{Time evolution of optical activity of chiral molecules in interaction with the environment at the strong-coupling limit for $k=0.1\Lambda$ and $\delta=0.1k$.}
  \label{Fig7}
\end{figure}

\begin{figure}[H]
  \centering
  \includegraphics[width=0.45\textwidth]{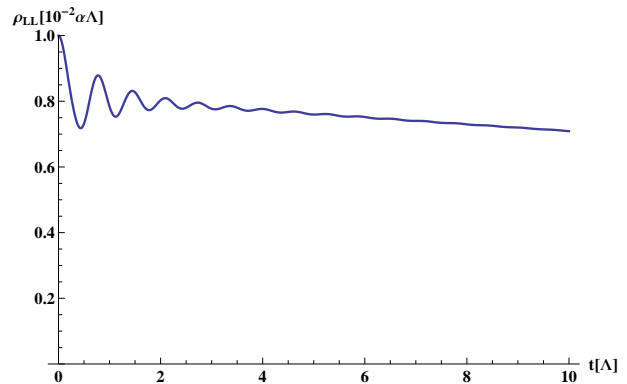}
  \caption{Time evolution of optical activity of chiral molecules in interaction with the environment at the strong-coupling limit for $k=10\Lambda$, $\delta=10k$.}
  \label{Fig8}
\end{figure}

The second differential equation in (\ref{E37}) describes the dynamics of the population of the left-handed state $\rho_{LL}$. The population of the right-handed state is given by $\rho_{RR}=1-\rho_{LL}$. For a fixed $k$ and $\Lambda$, the population of the chiral states are not dependent on the proportion of tunneling to tilt. Nevertheless, the set of differential equations (\ref{E37}) can be solved numerically within two limits. For $k\ll\Lambda$, the bath correlation time is much smaller than the typical time scale of the system. Then, the memory effects do not affect the system dynamics. Therefore, the decay rates approach their Markovian values. Accordingly, the non-Markovian master equation (\ref{E33}) as plotted in FIG.~\ref{Fig7} gives the similar dynamics to the Markovian master equation (\ref{E11}).
\noindent For $k\gg\Lambda$, the memory effect of the environment in the system dynamics is strong. In this case, the decay rates are oscillating, leading to an oscillatory behavior in the populations as plotted in FIG.~\ref{Fig8}.
\section{4. Conclusion}
We studied the dynamics of an ensemble of chiral molecules, described by an asymmetric double-well potential, in interaction with a bath of bosonic particles. The asymmetry was incorporated as the overall measure of the chiral interactions. Conventionally, the decoherence of chiral superpositions were studied under tunneling or localization regimes. However, we used these mechanisms along with each other. We carefully analyzed all contained effects at weak- and strong-coupling limits. The resulting master equations in (\ref{E11}) and (\ref{E33}) were solved via a set of coupled differential equations in (\ref{E28}) and (\ref{E37}), characterizing optical activity of the ensemble. We showed that at high-temperature limit, the decoherence effects resulted from the chiral interactions can discriminate two chiral configurations, stabilized due to tunneling process (FIG.~\ref{Fig1}). Specially, at tilt-dominant limit, environmental effects block the molecules in the initial chiral configuration (FIG.~\ref{Fig3}). At the strong-coupling limit, due to the memory effects of the environment, decoherence process cannot completely suppress the quantum correlations of the ensemble. This is reflected by the oscillatory behaviour of optical activity, specifically at short times.
\section{Acknowledgments}
We would like to thank Dr. Mohammad Bahrami for his instructive comments, and Dr. Mohammad Arjmand for an editorial reading of the article.

\end{document}